\documentclass{emulateapj}

\usepackage{times, graphicx}

\newcommand{\nad}{{Na~{\sc{i}}~D$_{1}$}}
\newcommand{\cak}{{Ca~{\sc{ii}}~K}}

\shortauthors{Grant et al.}

\begin{document}

\title{Wave Damping Observed in Upwardly Propagating Sausage-mode \\ Oscillations contained within a Magnetic Pore}
\author{S.D.T. Grant$^{1}$, D.B. Jess$^{1,2}$, M.G. Moreels$^{3}$,  R.J. Morton$^{4}$, D.J. Christian$^{2}$, I. Giagkiozis$^{5,6}$, G. Verth$^{5}$, \\ V. Fedun$^{7}$, P.H. Keys$^{1,5}$, T. Van Doorsselaere$^{3}$ and R. Erd\'{e}lyi$^{5,8}$}
\affil{$^{1}$ Astrophysics Research Centre, School of Mathematics and Physics, Queen's University Belfast, Belfast BT7 1NN, UK} 
\affil{$^{2}$ Department of Physics and Astronomy, California State University Northridge, Northridge, CA 91330, U.S.A.}
\affil{$^{3}$ Center for mathematical Plasma Astrophysics, Mathematics Department, KU Leuven, Celestijnenlaan 200B bus 2400, 3001 Leuven, Belgium}
\affil{$^{4}$ Mathematics and Information Sciences, Northumbria University, Newcastle Upon Tyne NE1 8ST, UK}
\affil{$^{5}$ Solar Physics and Space Plasma Research Centre (SP$^{2}$RC), The University of Sheffield, Hicks Building, Hounsfield Road, Sheffield S3 7RH, UK}
\affil{$^{6}$ Complex Optimization and Decision Making Laboratory, Automatic Control and Systems Engineering Department, University of Sheffield, Sheffield, S1 3JD, UK}
\affil{$^{7}$ Space Systems Laboratory, Department of Automatic Control and Systems Engineering, University of Sheffield, Sheffield S1 3JD, UK}
\affil{$^{8}$ Debrecen Heliophysical Observatory (DHO), Research Centre for Astronomy and Earth Sciences, Hungarian Academy of Sciences, 4010 Debrecen, P.O. Box 30, Hungary}

\email{sgrant19@qub.ac.uk}

\begin{abstract}
We present observational evidence of compressible magnetohydrodynamic wave modes propagating from the solar photosphere through to the base of the transition region in a solar magnetic pore. High cadence images were obtained simultaneously across four wavelength bands using the Dunn Solar Telescope. Employing Fourier and wavelet techniques, sausage-mode oscillations displaying significant power were detected in both intensity and area fluctuations. The intensity and area fluctuations exhibit a range of periods from $181-412$~s, with an average period $\sim$$290$~s, consistent with the global $p$-mode spectrum. Intensity and area oscillations present in adjacent bandpasses were found to be out-of-phase with one another, displaying phase angles of $6.12{\degr}$, $5.82{\degr}$ and $15.97{\degr}$ between 4170{\,}{\AA} continuum -- G-band, G-band -- {\nad} and {\nad} -- {\cak} heights, respectively, reiterating the presence of upwardly-propagating sausage-mode waves. A phase relationship of $\sim$$0{\degr}$ between same-bandpass emission and area perturbations of the pore best categorises the waves as belonging to the `slow' regime of a dispersion diagram. Theoretical calculations reveal that the waves are surface modes, with initial photospheric energies in excess of $35{\,}000$~W{\,}m$^{-2}$. The wave energetics indicate a substantial decrease in energy with atmospheric height, confirming that magnetic pores are able to transport waves that exhibit appreciable energy damping, which may release considerable energy into the local chromospheric plasma.
\end{abstract}


\section{Introduction}
An explanation of how the solar corona exhibits extra-ordinary temperatures exceeding $1$ MK remains a key goal of solar research \citep{Erdelyi2004, Klim2006, Tar2009}. Of greater importance is how chromospheric plasma can maintain temperatures of up to $10{\,}000$~K, since heating this region demands an even greater energy input to balance the extreme radiative losses \citep{With1977}. Atmospheric waves have long been proposed as the mechanism by which energy is transported from the solar surface to the upper atmosphere in order to sustain its heightened temperatures \citep{Schwarz1948}. The first verifiable evidence for oscillations in the solar atmosphere was documented by  \citet{Leighton1960}, and in the years since this discovery, observations of global and localised wave activity have become more definitive as a result of much improved instrumentation and processing techniques \citep[see, e.g., the recent reviews by][]{Ban2007, Math2013, Jess2015}. Alongside these observational improvements, our understanding of solar oscillations has increased through theoretical developments in the field of magnetohydrodynamics \citep[MHD; e.g.,][]{Roberts2000, Naka2005, Erdelyi2008}. From this, the nature of the Sun's atmosphere can be simplified as a region of magnetised plasma that can be manipulated through a number of physical mechanisms, including the ubiquitous convective flows present in the photosphere. MHD theory allows for an understanding of multiple wave modes in structured media, including sausage and kink modes, generated in the magnetically active regions of the solar atmosphere. The study of these waves in terms of their energetics and propagation through the solar plasma is vital in assessing the role oscillations play in atmospheric heating, given that the omnipresent solar acoustic (i.e., non-magnetic) waves have demonstrated insufficient energies to be the sole means of chromospheric heating \citep{Fossum2005, Bello2010}.

The various MHD wave modes that have been observed throughout the solar atmosphere can largely be categorised in terms of their compressibility. Weakly compressible MHD waves are distinguished by their inability to cause significant density perturbations within the plasma. Such modes identified thus far are the kink wave, which has been identified both in the chromosphere \citep[e.g.,][]{Kuk2006, Zaq2007} and the corona \citep[e.g.,][to name but a few]{Aschwanden1999, Nak99, Erdelyi2008b, VD2008}, alongside the more elusive and completely incompressible torsional Alfv\'{e}n wave \citep{Jess2009}. A consequence of the nearly incompressible nature of kink waves is their resistance to energy dissipation unless large Alfv\'{e}n speed gradients exist to help induce phase mixing \citep{Heyvaerts1983}, or provide mode conversion due to resonant coupling \citep{Sak1991}, making them the widespread focus of recent coronal heating studies. However, one drawback is that the proposed dissipation mechanisms for such waves exist on spatial scales below current observational resolution limits. As a result, there has only been tentative observational signatures of fast kink mode wave damping in the chromosphere \citep{He2009b, Kuridze2013, Morton2014b}, though the mechanism remains largely undetermined and it is uncertain whether these waves contribute directly to chromospheric heating.

Compressible wave modes are defined observationally by their ability to perturb the local plasma density. The greater capacity for mechanisms to extract energy from these waves, when compared to incompressible oscillations, continues to highlight their potential importance in providing heat energy to the outer solar atmosphere. The most observed compressible modes are the magneto-acoustic oscillations found in highly magnetic regions, such as sunspots, pores and magnetic bright points. Inclined magnetic fields allow global solar acoustic waves to propagate further into the atmosphere due to a weakened net gravitational force \citep{Bel1977}. Magneto-acoustic waves are thought to be associated with other observed oscillations in the solar atmosphere. For example, running penumbral waves are observed to propagate outwards along the penumbral magnetic fields surrounding sunspots \citep[e.g.,][]{Zirin1972}. They are thought to be the chromospheric manifestation of photospheric magneto-acoustic waves as they propagate upwards \citep*{Bloomfield2007, Jess13}, and have been shown to propagate upwards from the chromosphere with sufficient energy flux to contribute to the localised heating of coronal plasma \citep{Freij2014}. Compressible waves, in particular, are desirable wave modes given that many forms of energy dissipation have been identified, including resonant absorption \citep{goossens2001}, turbulent mixing \citep{vanBallegooijen2011}, viscosity \citep{Brag1965}, mode conversion \citep{Ulm1991} and thermal conduction \citep{Ofman2002}. 

Compressible MHD sausage modes are also of interest to the chromospheric/coronal heating debate. These waves are typically identified by periodic fluctuations in the intensity and/or area of magnetic structures such as pores, spicules or coronal loops \citep{Ed1983}. Given the high spatial resolution required to observe such fractional changes in area, early studies of sausage mode waves were limited to larger-scale coronal loops \citep[e.g.,][]{Aschwanden2004}, and it was only in recent years before the first lower atmospheric detection of sausage mode waves in whitelight images of pores was uncovered \citep*{Dorotovic2008}. As observing instrumentation has developed in capability, the ability to study sausage mode properties at high spatial and temporal resolutions has been evident. \citet{Mor2011} provided the first in-depth study of sausage modes in photospheric pores. The identified area oscillations often did not have simultaneous intensity changes, indicating these waves did not possess large-amplitude wave power provided there was no twist in the magnetic field. However, these waves were ascertained to be fast mode waves \citep{Mor2013b} and their observed periods indicated that they may be driven by the global $p$-mode oscillations, in contrast to the waves observed by \citet{Dorotovic2008}, whose longer periods were postulated as evidence of the presence of a magneto-acoustic gravity mode. Further photospheric analysis was conducted by \citet{Dorotovic2014}, where the investigation of separate pore and sunspot features identified the presence of both fast and slow sausage mode waves. These results showed that sausage modes form at the solar surface in a variety of different configurations, and have also been seen in fluctuations of spectro-polarimetric data \citep{Fujimura2009}, indicating a range of methods for photospheric detection and analysis. As the study of these waves at photospheric heights has developed, chromospheric detections initially appeared to be more elusive. However, the work of \citet{Mor2012} showed that sausage mode waves were in fact ubiquitous in fine-scale chromospheric magnetic structures, such as fibrils and mottles. These waves contained energy flux on the order of $11{\,}000$~W{\,}m$^{-2}$ and exhibited characteristics of leaky modes, suggesting that they were readily able to dissipate energy to the surrounding plasma. Given that the identified sausage-mode energy flux was greater than that of the simultaneous kink modes also found in these structures, the work was important in validating their potential for energy transportation to the upper regions of the solar atmosphere \citep[e.g.,][]{Jess12c, Moreels2015a}. However, what remains to be verified is a link between the photospheric and chromospheric observations of sausage modes, showing that the waves generated at the solar surface are indeed propagating to higher atmospheric heights. 

Magnetic pores are the visible signature of concentrated magnetic fields that prohibit surface convection over a large volume. They are the smaller counterparts to sunspots, with diameters between 1--6~Mm, and weaker magnetic field strengths up to $\sim1700$~G \citep{Sob2003}. There is an associated drop in the measured intensity, with the mean values of a pore being on the order of $40$\% lower than the quiescent solar surface \citep{Verma2014}. Given their lack of penumbra, pores have often been considered (and modelled) as a simple single magnetic flux tube \citep[][]{Simon1970, Hurlburt2000}. However, observations of bright structures within pores, such as umbral dots \citep*{Sob99}, could be an indication that pores may exhibit a more dynamic magnetic configuration, such as that described by \citet{Parker1979}. Their relatively small size has previously made detailed observations and analyses difficult, but with the advent of high resolution imaging instruments, they are now within scientific scope. This is advantageous since their small size leads to lessened internal forces when compared to sunspots, thus they are likely to be more dynamic and respond to bulk external forces more readily.

Here, we present observations of upwardly propagating slow sausage mode waves in a solar magnetic pore using high resolution imaging and spectral imaging from the Dunn Solar Telescope. We also provide the first intensive analyses of sausage mode energies with atmospheric height, from their formation in the photosphere through to high chromospheric locations.

\section{Observations and Data Processing}

\begin{figure*}
\begin{center}
%\epsscale{0.75}
%\plotone{Figure_01.jpg}
\includegraphics[width=0.6\textwidth,clip=]{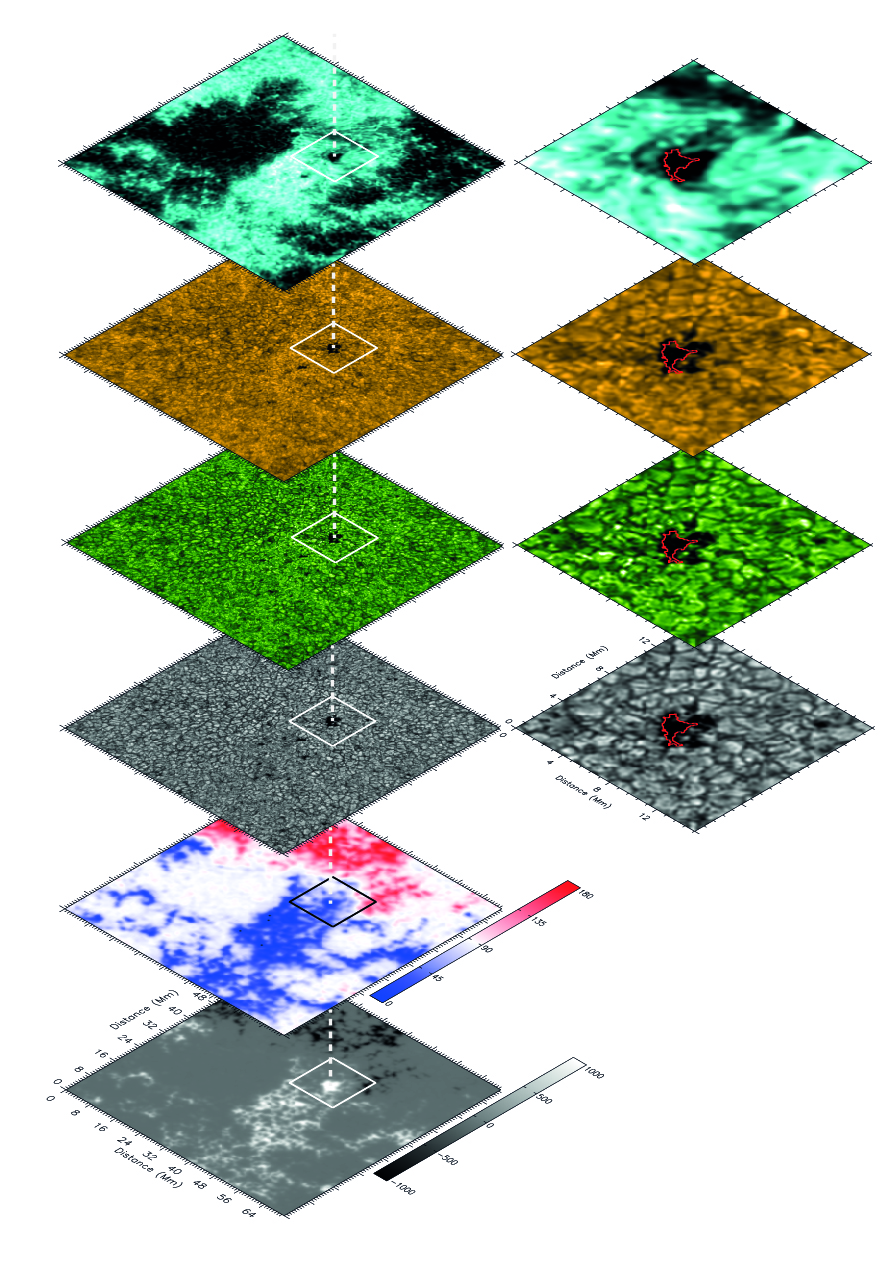}
\caption{The left column consists of co-spatial images of the full ROSA/CSUNcam field-of-view, stacked from the photosphere through to the chromosphere. From bottom to top, the images represent the vertical magnetic field strength, $B_{z}$, determined from HMI vector magnetograms with the scale saturated at $\pm$$1000${\,}G to aid clarity, magnetic field inclination angles, where $0^{\circ}$ and $180^{\circ}$ represent fields outwardly and inwardly normal to the solar surface, ROSA $4170${\,}{\AA} continuum, G-band, {\nad} and {\cak} images. A white dashed line interconnects the pore between bandpasses, while the solid boxes define the sub-fields displayed in the right-hand column. Here, the pore is shown in detail for each imaging bandpass, with the binary map pixels shown at each height using a red contour.}
\label{full}
\end{center}
\end{figure*}

The data presented here is an observational sequence obtained during $19$:$27-20$:$02$~UT on 2013 March 6 with the Dunn Solar Telescope (DST) at Sacramento Peak, New Mexico. The telescope was focussed on the active region NOAA $11683$, positioned at heliocentric co-ordinates ($113${\arcsec}, $-170${\arcsec}), or S$17.3$W$07.0$ in the conventional heliographic co-ordinate system, with seeing conditions remaining very good throughout the entire $35$~minute observing run. The Rapid Oscillations in the Solar Atmosphere \citep[ROSA;][]{Jess2010} camera system was used to image a $115\arcsec\times115\arcsec$ portion of the solar disk through continuum ($4170${\,}{\AA}; $52${\,}{\AA} FWHM), G-band ($4305.5${\,}{\AA}; $9.2${\,}{\AA} FWHM) and {\nad} core ($5895.9${\,}{\AA}; $0.17${\,}{\AA} FWHM) filters, each at a spatial sampling of $0{\,}.{\!\!}{\arcsec}12$ per pixel. A {\cak} core ($3933.7${\,}{\AA}; $1.0${\,}{\AA} FWHM) filter was also used in conjunction with a new camera addition to the DST's imaging suite. The California State University Northridge camera (CSUNcam) is an iXon X3 DU-897-EX\footnote{Full specifications available at http://www.andor.com.} model manufactured by Andor Technology, boasting a back-illuminated, $512$$\times$$512$~pixel$^{2}$ electron-multiplying CCD. The detector, triggering and readout architectures are identical to that of the existing Hydrogen-alpha Rapid Dynamics camera \citep[HARDcam;][]{Jess12a}, but incorporates an additional window coating designed to improve sensitivity in the blue portion of the optical spectrum, helping quantum efficiencies exceed $70${\%} at $4000${\,}{\AA}. To ensure the solar coverage of CSUNcam was similar to that of ROSA, a plate scale of $0{\,}.{\!\!}{\arcsec}230$ per pixel was chosen to provide a $118\arcsec\times118\arcsec$ field-of-view. 

In addition to ROSA and CSUNcam observations, the Interferometric BIdimensional Spectrometer \citep[IBIS;][]{Cav2006} was used to simultaneously sample the Ca II absorption profile at $8542.12${\,}{\AA} with a spatial sampling of $0{\,}.{\!\!}{\arcsec}097$ per pixel. IBIS employed $11$ discreet wavelength steps, each with $10$ exposures per wavelength to assist with image reconstruction, providing a complete scan cadence of $17.673$~s. A white-light camera, synchronised with the IBIS feed, was used to enable the processing and destretching of all narrowband images.    

To improve the clarity of the images, techniques of high-order adaptive optics \citep{Rim2004}, speckle reconstruction \citep*{Wol2008} and Fourier co-alignment \citep{Jess2007} were employed. Utilising $64 \rightarrow 1$ speckle restorations, the resulting cadences for continuum, G-band, {\nad} and {\cak} image sequences were $2.11${\,}s, $2.11${\,}s, $4.22${\,}s and $2.33${\,}s, respectively. The IBIS images were processed with a $10 \rightarrow 1$ restoration, and a blueshift correction was applied to account for the use of classical etalon mountings \citep{Cauzzi2008}. The {\nad} and {\cak} image sequences were then interpolated on to a constant $2.11${\,}s time grid to allow for easy comparisons between bandpasses, with the {\cak} images also spatially resampled to replicate the plate scale of the ROSA cameras. The end result is spatially and temporally coaligned image sequences spanning the full $35$~minute duration of the data set.

The Helioseismic and Magnetic Imager \citep[HMI;][]{Sch12} onboard the Solar Dynamics Observatory \citep*[SDO;][]{Pes12} was utilised to provide simultaneous vector magnetograms of active region NOAA~$11683$, in addition to contextual HMI continuum images for the purposes of co-aligning to the ground-based data sets. The Milne-Eddington vector magnetograms were obtained with a two-pixel spatial resolution of $1{\,}.{\!\!}{\arcsec}0$ and a cadence of $720$~s. The HMI data were processed using the standard {\tt{hmi\_prep}} {\sc{idl}} routine, which includes the removal of energetic particle hits. Next, $200${\arcsec} $\times$ $200${\arcsec} sub-fields were extracted from the processed images, with a central pointing approximately equal to that of the ground-based data. Using the HMI continuum image to define absolute solar co-ordinates, the ROSA and CSUNcam observations were subjected to Fourier-based cross-correlation techniques to provide sub-pixel inter-bandpass co-alignment. To do this, the plate scales of the ROSA and CSUNcam observations were first degraded to match that of the HMI continuum image\footnote{Data analysis was performed on full-resolution ({\rmfamily i.e.} non-degraded) image sequences.}. Then, calculations of squared mean absolute deviations were performed between the datasets, with the ground-based images shifted to best align with the HMI reference image. Following the implementation of co-alignment techniques, the maximum x- and y-displacements were both less than one tenth of an HMI pixel, or $0{\,}.{\!\!}{\arcsec}05$ ($\approx$$36$~km). Sample ROSA, CSUNcam and HMI images are displayed in Figure~\ref{full}.

\section{Analysis and Discussion}
Within our field-of-view there was a pronounced pore structure, as highlighted by the solid white boxes in Figure~\ref{full}, that lasted for the full duration of the data set. The pore was a well-defined dark discontinuity from the surrounding plasma. Thus, to isolate the pore for further study, we employed intensity thresholding techniques allowing the perimeter of the pore to be defined in relation to a decrease below the mean quiescent intensity. A quiet region, free from magnetic structures such as plage and bright points, was used to calculate the background mean intensity, $I_{\mathrm{ave}}$, and the resulting standard deviation, $\sigma$. Next, the pore was defined in each bandpass as pixels displaying intensity values less than ($I_{\mathrm{ave}} - N\sigma$), where $N$ is equal to $2.2$, $2.2$, $2.2$ and $2.1$ for the 4170{\,}{\AA} continuum, G-band, {\nad} and {\cak} bandpasses, respectively. While the value of $N$ may differ between adjacent bandpasses as a result of variable height-localised contrast ratios, once defined it remained fixed for each bandpass over the entire duration of the time series to ensure consistency when measuring time-dependent variables. 

By employing the intensity threshold values, pore time series for each bandpass were generated by isolating those pixels which lay at or below the specific threshold value. This created a $988$ ($35$~minute duration with a $2.11$~s cadence) image sequence for each bandpass with all solar structures except for the pore masked out. Then, the pore areas were time-averaged in each bandpass to establish the mean pore coverage at individual atmospheric heights. The time-averaged area of the pore defined by the intensity thresholds corresponded to $7.02 \pm 0.11${\,}Mm$^{2}$, $7.18 \pm 0.11 ${\,}Mm$^{2}$, $7.99 \pm 0.13${\,}Mm$^{2}$ and $8.12 \pm 0.19${\,}Mm$^{2}$ for the 4170{\,}{\AA} continuum, G-band, {\nad} and {\cak} bandpasses, respectively, indicating that the pore presented here expands minimally with height ($\sim$$15$\% area change between continuum and {\cak} layers). This expansion with height is expected due to the decrease in plasma pressure, although the fractional change in area is smaller than typically expected \citep[e.g.,][]{Solanki1999}. This may be a direct consequence of the strong magnetic fields present in the region of the pore at photospheric heights (left column of Figure~\ref{full}), which prevent rapid expansion as it extends upwards from the continuum level, or an unresolved twist in the field lines that would result in reduced expansion. With the pore perimeter well defined in each bandpass, it became possible to investigate and compare propagating and height-localised fluctuations in intensity and area, as will be discussed in the following subsections. 

\begin{figure*}
\begin{center}
%\epsscale{0.96}
%\plotone{Figure_02.eps}
\includegraphics[width=0.87\textwidth,clip=,viewport=0cm 0cm 65cm 45cm]{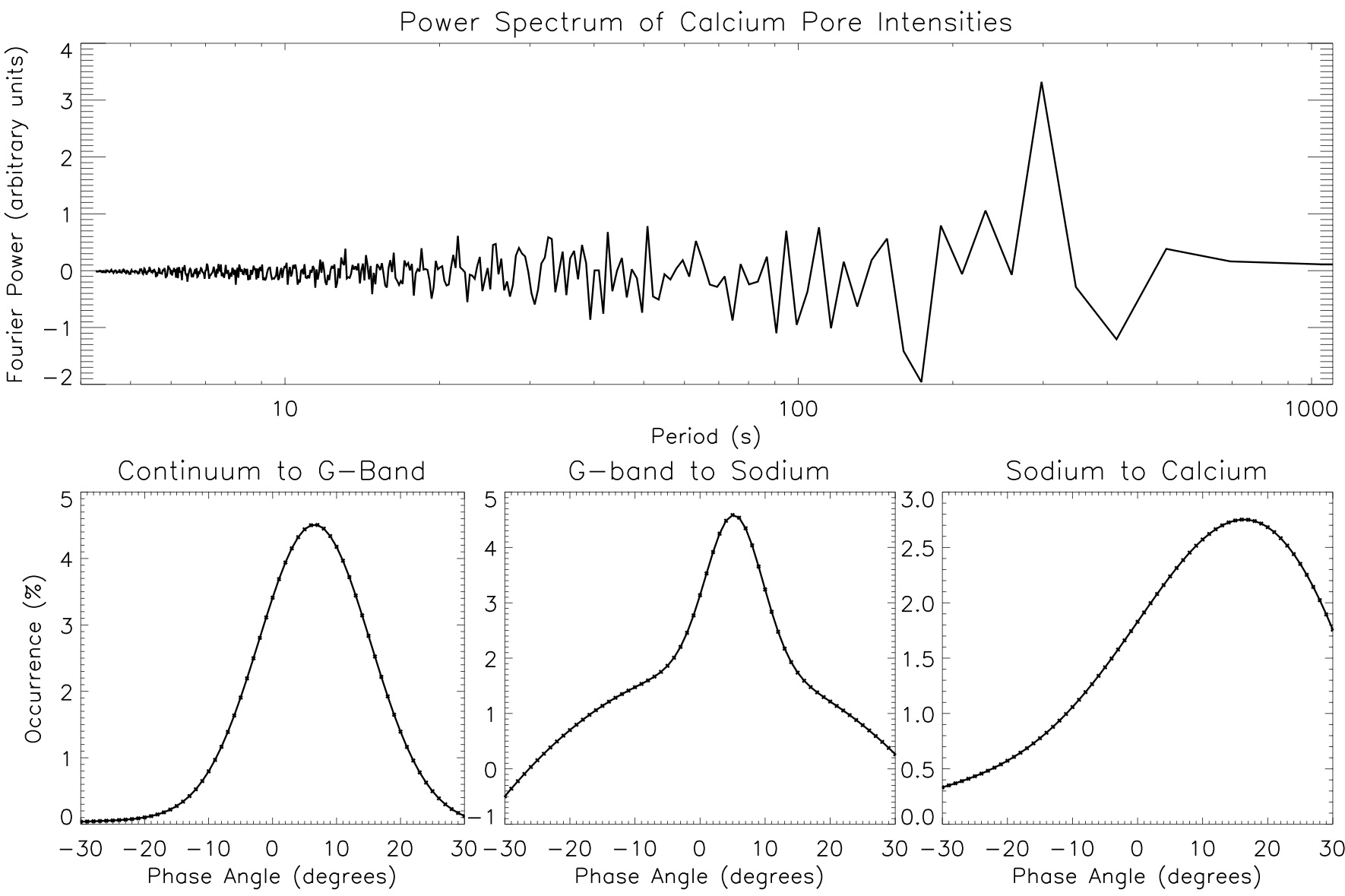}
\caption{The top panel is the sum of individual Fourier power spectra derived within the pore structure at {\cak} wavelengths. The Fourier power is displayed in arbitrary units, while the oscillatory periods are plotted on a logarithmic scale to better display oscillations synonymous with the longer-period $p$-mode spectrum. The lower panels are histograms of the phase angle values from the pixel-by-pixel intensity analyses for the 4170{\,}{\AA} continuum -- G-band, G-band -- {\nad} and {\nad} -- {\cak} bandpasses. For each inter-bandpass gap, a peak can be defined representing the most prominent phase angle observed.}
\label{histogram}
\end{center}
\end{figure*}

\subsection{Oscillations in the Pore Intensity}
\label{int}
To investigate whether the pore structure contained standing and/or propagating magneto-acoustic wave trains we studied co-spatial intensity fluctuations between adjacent bandpasses using the wavelet analysis techniques described by \citet{Jess12b}, and first introduced by \citet{Torr1998}. First, however, it was important to isolate pixels that remained part of the pore structure at all points in time {\it{and}} height. Only with these pixels determined can wavetrains that propagate through all available bandpasses be identified and statistically studied. Given that the 4170{\,}{\AA} continuum channel represented the smallest spatial coverage of the pore, it was used to determine which pixels are continually contained within the pore boundaries. The binary map was formed by co-adding all 4170{\,}{\AA} continuum pore perimeter maps, where values of `1' indicated part of the pore structure and values of `0' represented external non-pore areas, normalising by the image count ($988$), then setting all pixels with normalised intensities $<$$1$ to zero. Thus, only pixels that are contained within the pore perimeter for all steps in time are assigned a value of `$1$' in the final binary map, as indicated by the red contours in the right-hand panels of Figure~{\ref{full}}. The pixels contained within the 4170{\,}{\AA} continuum binary map were then compared to the time series for other bandpasses to verify their omnipresence throughout time and atmospheric height. The completed binary map contained $206$ non-zero pixels, and each pore image sequence was subsequently multiplied by the binary map to form a data set that could be compared directly between all atmospheric heights and time steps. 

Intensity time series for the 4170{\,}{\AA} continuum and G-band data sets were extracted for those pixels which lay within the confines of the binary map. The time series were then detrended by a first-order polynomial to remove long term variations in intensity and/or light levels, and normalised to their subsequent mean. Wavelet analysis techniques were applied to the time series to identify the presence of periodic signals. Strong oscillatory power was detected in both bandpasses, with periodicities predominantly in the range of $3$ to $5$ minutes, which overlaps well with the typical solar $p$-mode spectrum \citep{Lites1982}. Next, the phase difference analysis techniques detailed by \citet{Jess12b} were employed to investigate whether the oscillations were detectable co-spatially in the neighbouring bandpasses, and whether there was an inherent phase shift. Here, a positive phase angle indicates that a wave is first observed at a lower atmospheric height (i.e., the wave is propagating upwards from the 4170{\,}{\AA} continuum to the G-band), with criteria defined to ensure confidence in the final results. Firstly, only waves that exhibited a cross-correlation coefficient greater than $50$\% between the bandpasses were considered in an attempt to ensure the oscillatory signature corresponded to the same wave. Secondly, only waves which displayed a lifetime greater than $\sqrt{2}$$P$ in each bandpass were included, where $P$ is the period of the oscillation in seconds, to ensure these signatures are periodic rather than lone spikes in the time series. These rigorous criteria were implemented on each extracted time series, with approximately $1700$ suitable periodicities identified in the period range $210-412$~s, with a mean value of $290 \pm 31$~s. The derived individual phase angles between these detected oscillations were collected into a histogram employing $1{\degr}$ bins (as detailed in Figure~\ref{histogram}). A best-fit Gaussian was applied, with the centroided value identifying the most commonly occurring phase angle. In the case of 4170{\,}{\AA} continuum -- G-band, the dominant angle found was $\approx$$6.12 \pm 4.6 {\degr}$, indicating upward propagation between the formation heights of the 4170{\,}{\AA} continuum and G-band filtergrams. Subsequently, the same methodology was applied to the G-band -- {\nad} and {\nad} -- {\cak} time series. In the case of G-band -- {\nad}, periodicities in the range of $210-412$~s were identified (similar to that of the 4170{\,}{\AA} continuum -- G-band), with a prominent phase angle of $\approx$$5.82 \pm 3.62 {\degr}$. The {\nad} -- {\cak} output displayed a small decrease in the range of detected periodicities, with periods in the range of $181-362$~s being found to display a dominant phase angle $\approx$$15.97 \pm 8.73 {\degr}$. The upper panel of Figure \ref{histogram} displays a Fourier transform of {\cak} intensities synonymous with all pixels contained within the pore structure, here demonstrating significant power at periodicities in the range of $200$ - $300$~s. These relationships indicate the prevalence of upwardly propagating waves within the pore.

From inspection of Figure~\ref{histogram}, the majority of the derived phase angles are positive, with $75$\% of pixels showing a positive phase between 4170{\,}{\AA} continuum -- G-band, $70$\% between G-band -- {\nad}, and $78$\% between {\nad} -- {\cak}. The percentage of upwardly propagating waves determined here compares well with the work of \citet{Jess12b}, which ascertained that $73$\% of waves propagating between the $4170${\,}{\AA} continuum and G-band channels within kiloGauss magnetic elements were directed upwards. The existence of downwardly propagating waves is to be expected, and ultimately may be a consequence of reflections occurring at the chromospheric and transition region boundaries. In addition, the formation heights of the {\nad} and {\cak} filtergrams are difficult to quantify, with contribution functions being comprised of emission generated over a range of heights and opacities. As such, the phase angles calculated between these bands will have an intrinsic degree of scatter as a result of non-fixed formations heights and varying opacities in the vicinity of the highly magnetic pore. However, even with such effects, each inter-bandpass jump still demonstrates in excess of $70$\% upwardly propagating wave signatures.

\begin{figure}
\begin{center}
%\epsscale{1.1}
%\plotone{figure_03.eps}
\includegraphics[width=0.5\textwidth,clip=,viewport= 2cm 0cm 40cm 27cm]{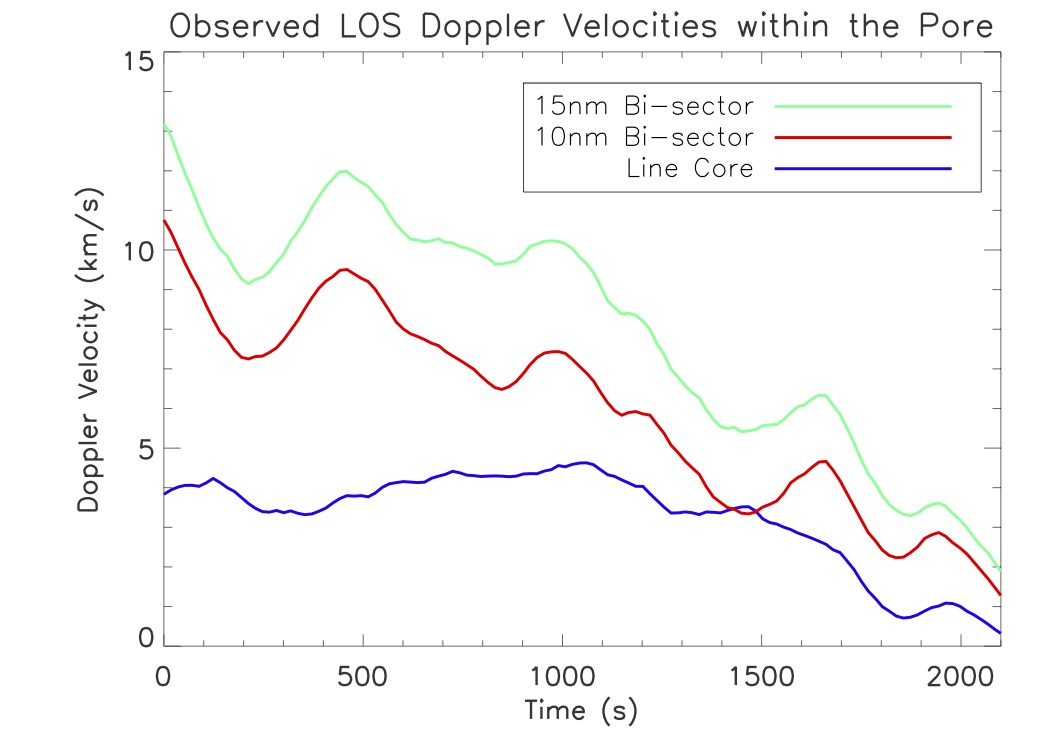}
\caption{The average line-of-sight velocities within the magnetic pore, measured from the IBIS $8542${\,}{\AA} Ca~{\sc{ii}} data. The blue line represents the velocity shifts in the line core, with the red and green lines denoting the velocity shifts at $100${\,}m{\AA} and $150${\,}m{\AA}, respectively, representing lower heights in the atmosphere away from the chromospheric line core. The line core average velocity is $3.27$~km{\,}s$^{-1}$, with the $100${\,}m{\AA} and $150${\,}m{\AA} bi-sectors exhibiting average velocities equal to $5.93$~km{\,}s$^{-1}$ and $8.01$~km{\,}s$^{-1}$, respectively.}
\label{velocities}
\end{center}
\end{figure}

The phase speed of the waves can be estimated by combining the observed phase angle alongside the formation heights of the 4170{\,}{\AA} continuum and G-band bandpasses derived by \citet{Jess12b}. While the average separation between formation heights was found to be $\sim$$75$~km, structures residing within more magnetic environments (such as intergranular lanes and magnetic bright points), were found to display larger separation heights on the order of $\sim$$100$~km. While more difficult to constrain, the formation heights of the {\nad} and {\cak} filtergrams can be estimated as $\sim$$400$~km \citep{Simon08} and $\sim$$800$~km \citep{Beebe1969}, respectively. Employing our periodicity and phase delay measurements from the 4170{\,}{\AA} continuum to G-band, along with the traversed distance extracted from \citet{Jess12b}, enables us to calculate the resulting phase velocities of the observed waves. Assuming the prominent phase angle of $6.12{\degr}$, found between 4170{\,}{\AA} continuum and G-band filtergrams, represents the phase velocity of the waves with a mean period of $290$~s, the wave travel time can then be calculated as $6.12\degr / 360\degr \times 290$~s $=4.92$~s. Thus, an average propagation distance of $\sim$$75$~km leads to an estimated wave speed of $\sim$$15$~km{\,}s$^{-1}$. At first glance, this value is in excess of what is expected for an isolated wave propagating in the solar photosphere \citep{Ed1983}. However, investigation of the line-of-sight Doppler and bi-sector velocities obtained with IBIS revealed a large bulk upflow within the confines of the pore. Figure~\ref{velocities} displays the spatially averaged Doppler and bi-sector velocities within the pore at various positions in the line profile. As can be seen from the bi-sector velocities, which best represent lower atmospheric heights, a global upflow with an average velocity of $\sim$$8$~km{\,}s$^{-1}$ is present, with a peak value approaching $14$~km{\,}s$^{-1}$ recorded. The observed decrease in the plasma upflow velocities towards the Ca~{\sc{ii}} line core remains consistent with the previous work of \citet{Cho2013}. With the addition of a bulk plasma upflow, the initial speed calculated can be considered as a Doppler-shifted phase speed, which incorporates both the phase velocity of the wave and the bulk motion of the transfer medium. Therefore, an estimate of the true phase speed of the waves can then be derived by subtracting the line-of-sight measurements from the derived velocity, thus producing a phase speed on the order of $5$~km{\,}s$^{-1}$ at lower atmospheric heights, which is similar to, if not below, the adiabatic sound speed predicted for these locations \citep[see, e.g.,][]{Jess12b}. Thus, the effect of the waves being superimposed on top of a simultaneous bulk upflow, alongside the potential effects of a twisted magnetic field configuration \citep[e.g.,][]{Erdelyi2008c} may account for the large measured Doppler-shifted phase velocity.

~

\subsection{Oscillations in the Pore Area}
\label{Area}
Upwardly propagating intensity oscillations may be a direct signature of magneto-acoustic wave modes \citep[e.g.,][]{DePontieu2004, Jess12a, Freij2014}. Therefore, in order to verify whether the detected oscillations also have the observational signatures of sausage mode waves, it was necessary to consider whether simultaneous periodic changes in the area of the pore existed. When compared to the pixel-by-pixel analysis employed for the intensity oscillations, which provided $206$ individual pixel measurements per time step, the examination of periodic area changes has statistical limitations due to only one area measurement being possible per instant in time (rather than 206 individual intensity measurements). Nevertheless, the extraction of phase delays between pore area sizes as a function of atmospheric height was performed identically to that described in Section~\ref{int}. 

The analyses identified area oscillations of similar period ranges ($181-412$~s) to that of the intensity oscillations, thus indicating the presence of compressional waves that are able to modify the cross-sectional area with time. The results of the phase analysis on neighbouring bandpasses again provided a consistently positive phase angle, reiterating the presence of upwardly propagating compressive waves. Specifically, the phase delays found in area oscillations between the 4170{\,}{\AA} continuum -- G-band, G-band -- {\nad} and {\nad} -- {\cak} bandpasses were $6.35 \pm 2.16{\degr}$, $5.72 \pm 6.7{\degr}$ and $16.76 \pm 13.3{\degr}$, respectively. These results are comparable to the peak values found for the pixel-by-pixel intensity oscillations (see Figure~{\ref{histogram}}). Thus, the identification of synchronous intensity and area oscillations, alongside the remarkable similarity between the phase delays for both sets of perturbations, provides evidence for the presence of upwardly propagating sausage mode waves within the pore.  

\subsection{Determining the Wave Mode}
\label{PhaseSameBand}
Employing the phase analysis techniques independently to both intensity and area fluctuations as a function of atmospheric height has confirmed the presence of upwardly propagating sausage mode oscillations. However, these oscillations may fall within the `slow' or `fast' sections of the dispersion relation. The categorisation of waves in this manner allows limitations to be imposed upon properties such as the phase speed \citep{Ed1983, Evans1990, Erdelyi2010} and dissipative potential through mechanisms such as thermal conduction  \citep{VD2011}. Therefore, in order to categorise whether the observed sausage mode oscillations are `fast' or `slow', the phase relationships between localised intensity and Lagrangian area oscillations within each respective bandpass was investigated. Theory predicts an in-phase relationship between intensity and area for slow-mode waves, while a phase lag of $\sim$$180{\degr}$ suggests the presence of fast-mode oscillations \citep{Mor2013b}.

The criteria used in sections \ref{int} and \ref{Area} were implemented to calculate the relevant phase relationships for the four band passes. Prominent phase delays between intensity and area of $17 \pm 21{\degr}$, $19 \pm 16{\degr}$, $11 \pm 3{\degr}$ and $21 \pm 20{\degr}$ were found for the 4170{\,}{\AA} continuum, G-band, {\nad} and {\cak} bandpasses, respectively, with the area consistently leading the intensity fluctuations in each bandpass. In accordance with \citet{Mor2013b}, this suggests a dominant presence of slow mode waves.

{Sausage mode waves have already been shown to exhibit strong velocity components along the tube axis \citep{Fujimura2009}. As such, oscillatory patterns in line-of-sight (LOS) Doppler velocities can be employed as a diagnostic tool when comparing with simultaneous intensity fluctuations. Phase relationships between the LOS Doppler velocity and the Eulerian intensity are identified in \cite{Mor2013a}. It is shown that slow propagating modes will demonstrate in-phase behaviour. The LOS velocity can be determined from the Ca~{\sc{ii}} data from IBIS, where a series of velocity bi-sectors were calculated. Such bi-sector measurements sample the plasma velocities at locations in the profile wings (i.e., away from the chromospheric line core), and are therefore analogous to velocity measurements at lower heights in the solar atmosphere. Thus, the LOS values calculated from the wings of the Ca~{\sc{ii}} IBIS scans will be formed higher than the pure solar continuum (i.e., the 4170{\,}{\AA} ROSA continuum), yet lower than the chromospheric formation heights of the Ca~{\sc{ii}} line core, including those found in the {\cak} ROSA filtergrams. Therefore, in the case of the present data, the velocity bi-sectors are most compatible with the {\nad} observational time series. Oscillations found in the 150{\,}m{\AA} bi-sector velocities are observed to have a phase relationship of $\approx$$0.4 \pm 0.1{\degr}$ with the Eulerian intensity oscillations derived from the {\nad} filtergrams. 

The approximately in-phase relationship between the LOS Doppler velocities and the Eulerian intensities provides confidence in a slow mode interpretation. The slight discrepancies found between observation and theory in terms of the area and intensity phase angles (i.e., $17{\degr}$, $19{\degr}$, $11{\degr}$ and $21{\degr}$; see above) may be an indication that there is a superposition of both fast and slow sausage modes within the pore (albeit dominated by the slow modes). It may also be the result of varying opacity across the diameter of the pore, hence inducing variations in the pixel-by-pixel formation heights sampled by the filtergrams. This phase lag could also be caused by dissipative effects, as observed by \citet{VD2011b}. However, due to the clear in-phase relationship between the LOS Doppler velocities and the Eulerian intensities, we are able to best interpret the observed waves as propagating slow axisymmetric modes.

\subsection{Assessing the Propagating Energy Flux of the Waves}
Using a newly developed framework for evaluating the energy flux of axisymmetric waves \citep{Moreels2015b}, a detailed analysis of the energy of the observed MHD waves at each atmospheric height can be deduced for the first time. Before the energy analyses can be undertaken, features of both the pore, its surroundings and individual sausage oscillations must be estimated as inputs to the energy calculations.

\begin{figure*}
\begin{center}
%\epsscale{0.85}
%\plotone{Figure_04.eps}
\includegraphics[width=0.8\textwidth,clip=,viewport=0cm 0cm 50cm 38cm]{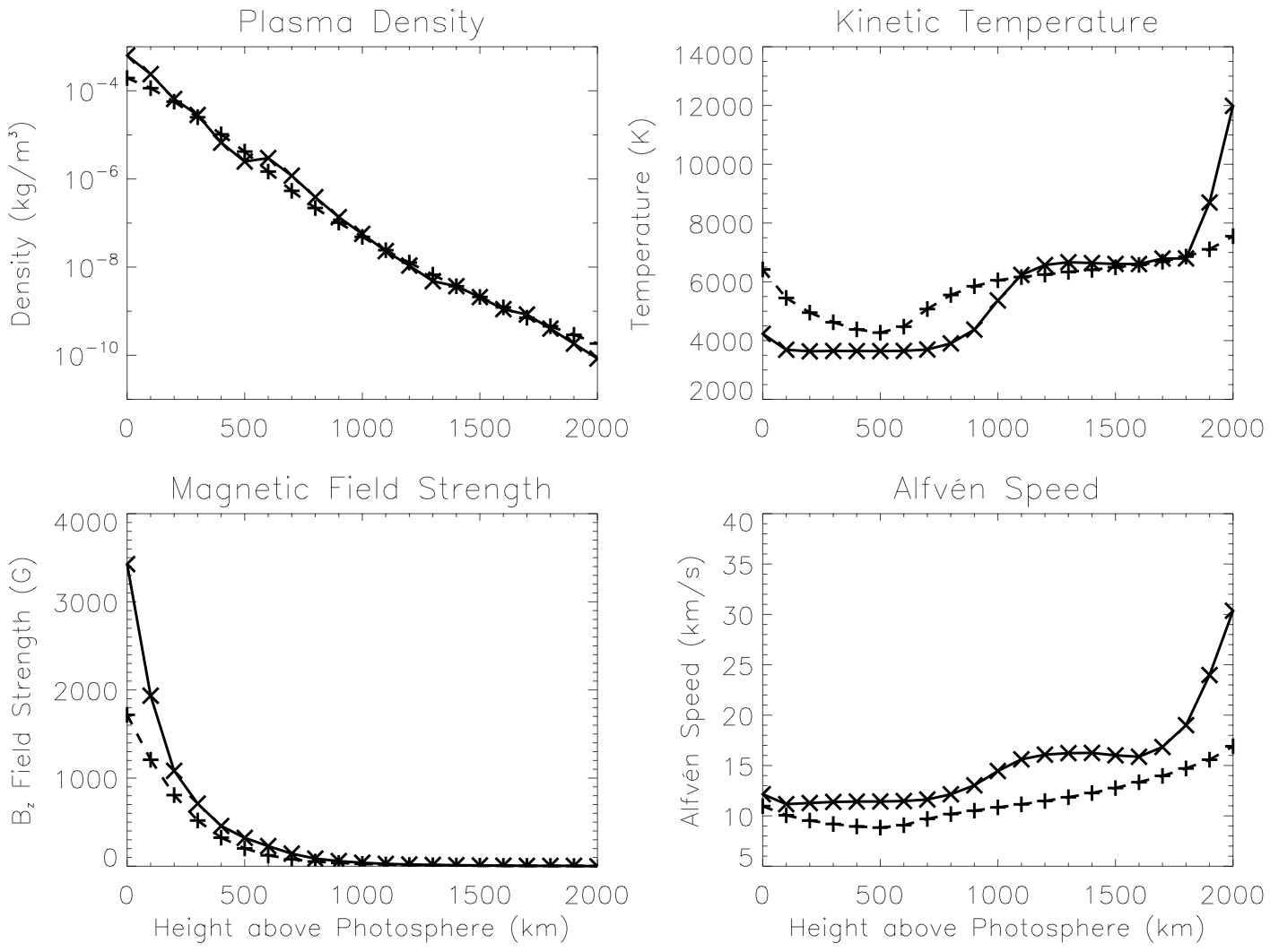}
\caption{The plasma density (upper-left), kinetic temperature (upper-right), vertical magnetic field strength ($B_{z}$; lower-left) and derived Alfv{\'{e}}n speed (lower-right), estimated as a function of height above the solar photosphere for internal \citep[solid line; deduced by scaling the `hot' umbral model of][]{Mal1986} and external \citep[dashed line; estimated directly from the VAL--D model of][]{Ver81} pore configurations.}
\label{pore_parameters}
\end{center}
\end{figure*}

\subsubsection{Deriving Plasma Parameters}
\label{Parameters}
In order to investigate the energy content of the waves, local plasma parameters must be identified that can then be integrated into a similar model atmosphere. Employing the Very Fast Inversion of the Stokes Vector \citep[VFISV;][]{Bor11} algorithm, the line-of-sight magnetograms provided by the HMI instrument were decomposed to define a magnetic component perpendicular the solar surface (i.e., $B_{z}$). The formation height of the HMI magnetograms corresponds to $\sim$$300${\,}km above the photosphere \citep{Nor06, Fle11}, with the pore revealing itself as a positive polarity with an area-averaged vertical magnetic field strength $\approx$$900${\,}G, peaking at nearly $1500${\,}G in the central region of the pore (lower-left panel in Figure ~\ref{full}). Next, the \citet{Mal1986} `hot' umbral model `L' was scaled so the magnetic field strength of the model (at a height of $\sim$$300${\,}km) matched that of our HMI observations. The `hot' umbral model was chosen as the best approximation to the pore, given that it is significantly smaller than a typical sunspot structure, and as a result is likely to be marginally hotter at its core due to the fact that convective processes will not be as severely inhibited \citep{Sob99}. Utilising the (non-scaled) \citet{Mal1986} density values, alongside the scaled magnetic-field strengths, enabled the calculation of the internal Alfv{\'{e}}n speed, $V_{A}$, as a function of atmospheric height using the relation, 
\begin{equation}
V_{A} = \frac{B_{z}}{\sqrt{\mu_{0}\rho}} \ ,
\end{equation}
where $B$ is the magnetic field strength, $\mu_{0}$ is the magnetic permeability and $\rho$ is the density of the local plasma. The solid black lines in Figure~\ref{pore_parameters} displays the plasma densities, kinetic temperatures, scaled magnetic field strengths and Alfv{\'{e}n speeds as a function of atmospheric height. It is clear that for lower atmospheric heights up to $\sim$$700${\,}km, the Alfv{\'{e}}n speed determined is $\sim$$12${\,}km{\,}s$^{-1}$.

It was also necessary to employ model atmosphere parameters, corresponding to solar plasma conditions outside the magnetic pore, that could be directly compared with the scaled \citet{Mal1986} values. Examination of the bottom-left panel of Figure~{\ref{full}} reveals that the pore exterior is still magnetic in nature, with field strengths of the order of a few hundred Gauss. As a result, the \citet*{Ver81} model `D' atmosphere was chosen to best represent the exterior of the pore structure. The VAL-D parameters are consistent with magnetic network elements, and thus best represent the nature of the solar plasma immediately outside the pore structure. However, as the atmospheric height scales differ between the \citet{Ver81} and \citet{Mal1986} models, each set of parameter measurements were interpolated on to a constant height separation grid equal to $100${\,}km. The resulting VAL-D plasma densities, kinetic temperatures, magnetic field strengths and derived Alfv{\'{e}}n speeds are displayed in Figure~{\ref{pore_parameters}} using a dashed line. Of notable comparison to the scaled \citet{Mal1986} values are the temperatures, magnetic field strengths and Alfv{\'{e}}n speeds, all of which are smaller within the lower solar atmosphere, thus remaining consistent with our ROSA/CSUNcam imaging and HMI magnetogram observations. 

Although these models provide one of the best representations of the pore and its surroundings currently available, their accuracy of the true conditions observed may be limited. In particular, the modelled magnetic field is scaled through observations at a single isolated height, leading to potential uncertainties as a result of complex magnetic structuring. This may produce an over-estimation of magnetic field strength in the lower atmosphere, especially considering our derived field strength at the solar surface is on the order of $3000$~G. Furthermore, a typical sunspot model will incorporate atmospheric expansion greater than the $15$\% we observed for the pore. As a result, this may lead to the modelled pore exhibiting greater plasma densities at higher atmospheric heights compared to real life, and as such, the internal Alfv{\'{e}}n speed calculated here may be large in relation to typical chromospheric channels. The lack of measured expansion may also lead to larger longitudinal magnetic field strengths at chromospheric heights than predicted by the model atmosphere as a result of magnetic flux conservation. However, given that the pore structure is not observable at higher heights in simultaneous SDO imaging, this suggests that the magnetic field may have fanned out into the surrounding magnetic canopy, as one would expect for larger sunspot structures at chromospheric and coronal heights. As a consequence, such fanning will lead to a progressive decrease in the longitudinal magnetic field strength with atmospheric height, in agreement with the Maltby model (see the lower-left panel of Figure~\ref{pore_parameters}). In future, additional observational steps can be taken to mitigate the issues highlighed here. In particular, complimentary spectro-polarimetric observations, using absorption lines such as Fe~{\sc{i}}~$6302${\,}{\AA}, would allow for an accurate measurement of the magnetic field at a different atmospheric height, hence better constraining the magnitudes of the magnetic fields. While we openly discuss any potential limitations of the model parameters derived here, they still remain one of the best approximations currently available.

\begin{deluxetable*}{ l c c c c c c c c }
\tablecaption{Properties of the upwardly propagating 210s and 290s oscillations}
%\rotate
\tabletypesize{\scriptsize}
\tablehead{ Parameters & \multicolumn{2} {c}{Continuum} & \multicolumn{2} {c} {G-Band} & \multicolumn{2} {c} {\nad} &\multicolumn{2} {c} {\cak} }
\startdata
  & 210s & 290s & 210s & 290s & 210s & 290s & 210s & 290s \\ 
Intensity Perturbation (\%) & 3.6 $\pm$ 2.5 & 3.4 $\pm$ 2.4 & 3.6 $\pm$ 2.5 & 2.8 $\pm$ 1.9 & 2.95 $\pm$ 2.09 & 2.2 $\pm$ 1.5 & 5.53 $\pm$ 3.91 & 5.6 $\pm$ 4 \\ 
Area Perturbation (\%) & 1.53 $\pm$ 1.08 & 2.1$\pm$ 1.48 & 2.92 $\pm$ 2.07 & 2.1 $\pm$ 1.48 & 3.25 $\pm$ 2.29 & 3.25 $\pm$ 1.62 & 4.75 $\pm$ 3.36 & 5.8 $\pm$ 4.1\\
Phase Speed (km/s) & 4.03 $\pm$ 0.66 & 2.5 $\pm$ 1 & 2.26 $\pm$ 0.83 & 1.9 $\pm$ 0.68 & 2.09 $\pm$ 0.89 & 2.08 $\pm$ 0.9 & 3 $\pm$ 1.5 & 2.68 $\pm$ 1.3 \\ 
Energy Flux (kW/m$^{2}$) & 54.5 $\pm$ 26.3 & 38.4 $\pm$ 21.4 & 70.1 $\pm$ 35.8 & 34.1 $\pm$ 13.5 & 1.99 $\pm$ 0.8 & 2.48 $\pm$ 1.03 & 0.12 $\pm$ 0.08 & 0.13 $\pm$ 0.08 \\ 
\enddata
\label{Periods}
\end{deluxetable*}

\subsubsection{Identifying and Characterising Individual Perturbations}
\label{waveperturb}

Thus far, the axisymmetric oscillations have been considered as a superposition of multiple individual periodicities. However, in order to implement the energy analysis of \citet{Moreels2015b}, the oscillatory properties of individual periodicities must be extracted from the time series. It was therefore necessary to isolate linear oscillatory perturbations in order for the energy analysis to be valid, but also to isolate waves that could be identified over a narrow period range in order to limit the associated uncertainties in the frequency domain. Periodicities were investigated by filtering the intensity and area time series in the frequency domain through use of Fourier transforms. Two periods in particular were identified, a $210$~s sausage mode that was present for the first $844$~s of the observations, and a $290$~s wave that was present for the entire $35$~min duration of the time series (see, e.g., the upper panel of Figure~{\ref{histogram}}). Both wave modes exhibited clear sinusoidal behaviour, allowing area and intensity amplitudes to be easily extracted as a percentage fluctuation above the mean, as detailed in Table~1. 

An important input to the energy calculations is the phase speed of the waves as they propagate upwards, and the methods of \citet{Moreels2015a} were used to calculate the phase speed at each atmospheric height. The methodology utilises a uniform flux tube as an equilibrium model, and combined with linear MHD theory, establishes a link between the phase speed of an oscillation and the amplitude of the total intensity and area fluctuations. A dimensionless amplitude ratio, $A$ is defined as,
\begin{equation}
A = \frac{\delta I / I_{0}}{\delta S / S_{0}} \ ,
\end{equation}
where $\delta I$ is the amplitude of the intensity perturbation, $I_{0}$ is the mean intensity, $\delta S$ is the amplitude of the area perturbation and $S_{0}$ is the mean area. The phase speed is therefore defined as, 
\begin{equation}\label{eq:seismology}
\frac{\omega}{k} = c_{S} \sqrt{\frac{\pm A - 1}{\pm A - 1 + (\gamma - 1)(h\nu / k_{B}T)}} \ ,
\end{equation}
where $\omega/k$ is the phase speed, $\gamma$ is the ratio of specific heats, $h$ is the Planck constant, $\nu$ is the
frequency at which the observation was taken, $k_{B}$ is the Boltzmann constant, and $T$ is temperature within the pore. The sound speed, $c_{S}$, is defined as 
\begin{equation}\label{eq:sound}
c_{S} = \sqrt{\frac{\gamma R T}{\mu}} \ ,
\end{equation}
where  ${\gamma}$ is the ratio of specific heats, {\it{R}} is the gas constant, {\it{T}} is the temperature and $\mu$ is the mean molecular weight. However, equation \ref{eq:seismology} is, strictly speaking, only valid for continuum emission and not for specific emission/absorption lines. Therefore equation \ref{eq:seismology} should only be applied to the continua found in the 4170{\,}{\AA}, G-band, {\nad} and {\cak} filters. While significant numbers of absorption features are present in both the 4170{\,}{\AA} and G-band filter response curves, the large spectral coverage of the filters means that they will predominantly be comprised of solar continua. For the {\nad} and {\cak} image sequences we need a different inversion formula. Let us assume that for any optically thick absorption line we can write the intensity in terms of the local density and temperature, i.e., $I=F(\rho,T)$ where $F$ is a smooth function. Following the same methodology as \citet{Mor2013b, Moreels2015a} we find,
\begin{equation}\label{eq:seismology2}
 \frac{\omega}{k} = c_S\sqrt{\frac{\pm A - 1}{\pm A - 1 + \left[\frac{\rho_0}{F_0} \frac{\partial F}{\partial\rho}+(\gamma-1)\frac{T_0}{F_0} \frac{\partial F}{\partial T}\right]}},
\end{equation}
%\begin{equation}\label{eq:seismology2}
 %\frac{\omega}{k} = c_S\sqrt{\frac{\pm A - 1}{\pm A - 1 + \left[(\rho_0/F_0) (\partial F/\partial\rho)+(\gamma-1)(T_0/F_0) (\partial F/\partial T)\right]}} \ ,
%\end{equation}
where both of the derivatives are to be evaluated at $\rho_0$ and $T_0$ respectively. The indices `zero' stand for the equilibrium parameters, i.e., $F_0=F(\rho_0,T_0)$. Indeed, when we use $F(\rho,T)=(2 h {\nu^3} / {c^2}) \exp(-h\nu/k_{\mathrm{B}} T)$ we recover equation \ref{eq:seismology}.

Unfortunately, the use of equation \ref{eq:seismology2} requires a detailed knowledge of the precise formation heights of the {\nad} and {\cak} spectral lines. Examining other lower atmospheric spectral features, including the H$\alpha$ and Mg~{\sc{ii}} lines, \citet{Len2012, Len2013} demonstrated the difficulty in accurately linking the observed intensity to the localised density/temperature. The authors claim that due to the chromosphere not being in local thermodynamic equilibrium, it is therefore difficult to isolate and quantify the local thermal properties of the plasma. However, \citet{Len2013} revealed that the temperature at $\tau=1$ is often related to the radiation brightness temperature. Extending these results to {\nad} and {\cak} filtergrams, we are able to use equation~\ref{eq:seismology} to interpret out findings with $T$ acting as the radiation brightness temperature. An additional justification for using equation \ref{eq:seismology} comes from the fact that due to both filters having non-infinitely narrow spectral windows, they will naturally sample a considerable portion of the optical continuum. For example, the {\nad} and {\cak} filters cover $0.17${\,}{\AA} and $1${\,}{\AA} wide regions, respectively, with continua being clearly visible in the form of granulation in the image sequences.

The inferred phase speeds, which are detailed in Table~1, also provide further verification that the waves are slow modes. Slow sausage modes are shown to propagate below the sound speed, $c_{S}$, of the flux tube \citep{Ed1983}. The sound speed inside the pore is derived from equation \ref{eq:sound}, using the atmospheric parameters displayed in Figure~{\ref{pore_parameters}}, and was estimated as $\sim$$7$~km{\,}s$^{-1}$ at the photospheric layer, which is greater than the estimated phase speeds of the waves in each band pass, thus supporting a slow mode interpretation. The sound speed estimated in this way is also consistent with the numerically calculated value of the sound speed found in \citet{Jess12b}. Importantly, however, is the fact that the phase speeds determined for the atmospheric heights corresponding to the 4170{\,}{\AA} continuum and G-band are consistent with the phase speeds estimated from the time-lag measurements detailed in Section~{\ref{int}}. This self-consistency provides reassurance in our interpretation that the observed waves are upwardly propagating slow axisymmetric oscillations.

\subsubsection{Defining whether the Modes are Surface or Body}
Sausage mode waves can be further categorised based on their structuring within the flux tube. Body modes exhibit periodic structure along the tube radius for radial modes greater than zero, whereas surface modes do not \citep{Ed1983}. The distinction between surface and body modes has been shown to have an effect on the phase speed of the waves \citep{Mor2013b}, but of greater importance are the implications related to the calculated energy values. As detailed in \citet{Moreels2015b}, the additional internal oscillations associated with body modes demand that the two cases be considered separately. As such, individual energy equations for both surface and body modes are defined by \citet{Moreels2015b}. Therefore, it is necessary to verify whether the oscillations observed in the pore are surface or body modes before energy calculations can be undertaken.

A neat verification of the surface/body nature of sausage modes was initially presented by \citet{Mor2013b}, whereby the ratio of wave constants, $\kappa$, at a given height may be defined as,
\begin{equation}
\label{kappa}
\kappa^{2} = \frac{(k^{2}c_{S}^{2} - \omega^{2}) (k^{2} v_{A}^{2} - \omega^{2})}{(c_{S}^{2} + v_{A}^{2})(k^{2}c_{T}^{2} - \omega^{2})} \ ,
\end{equation}
where $P$ represents the period of an oscillation, $k = 2\pi / P v_{\mathrm{ph}}$ is the longitudinal wavenumber involving the phase speed of the wave, $v_{\mathrm{ph}}$, and $c_{T}$ is the tube speed of wave defined as,
\begin{equation}
c_{T} = \frac{c_{S} v_{A}}{(c_{S}^{2} + v_{A}^{2})^{2}} \ ,
\end{equation}
with all other terms defined in previous sections. It is stated in \citet{Mor2013b} that in the case of surface modes, the evaluated expression for $\kappa^{2}$ will be positive. All values needed to calculate $\kappa^{2}$ can be extracted from the model parameters displayed in Figure~\ref{pore_parameters}, or from those defined in Table~\ref{Periods}. Using these terms, $\kappa^{2}$ was evaluated to be positive over all atmospheric heights, indicating that the sausage oscillations observed in this pore are slow propagating surface modes.

\begin{figure}
\begin{center}
%\epsscale{1.2}
%\plotone{figure_05.eps}
\includegraphics[width=0.5\textwidth,clip=,viewport = 2cm 0cm 38cm 28cm]{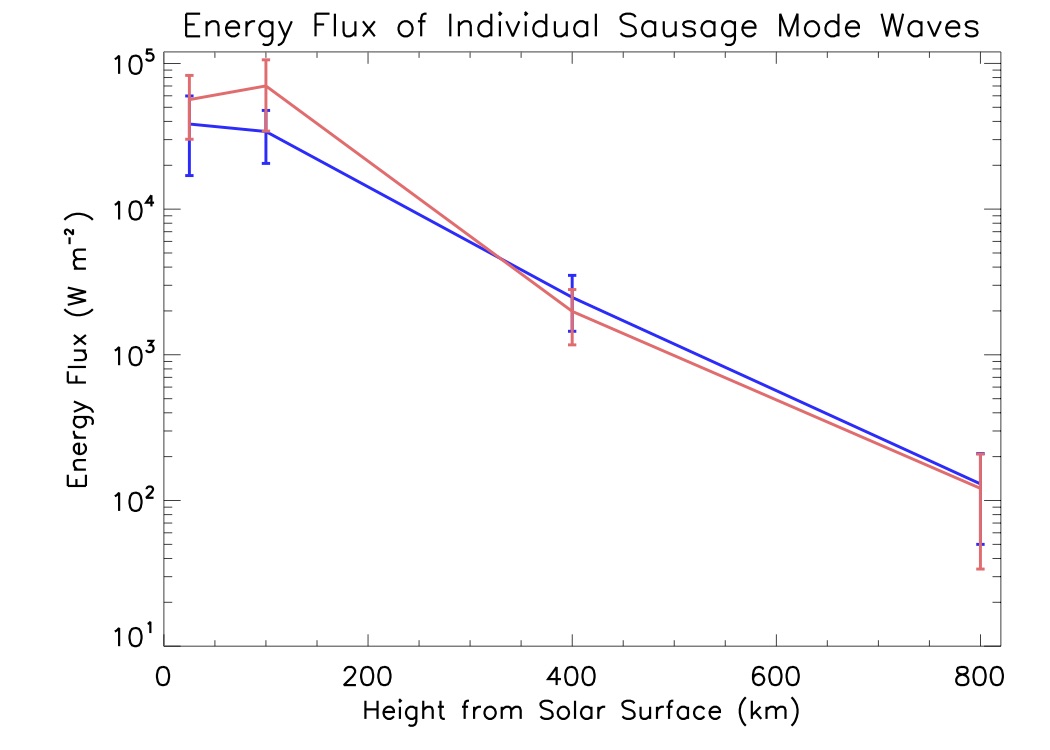}
\caption{The calculated energy flux of $210$~s (red) and $290$~s (blue) sausage mode oscillations, both in W{\,}m$^{-2}$. The heights from the solar surface are approximate values that are stated in section \ref{int}, with the energy flux values plotted on a logarithmic scale to better highlight the rapid decrease in observed energies with increasing atmospheric height.}
\label{energy}
\end{center}
\end{figure}

\subsubsection{Calculating the Energy Flux}
With the parameters of the plasma inside and outside the pore defined, alongside the phase speeds for individual periodicities at each height and the identification of their surface mode nature, the energy flux can subsequently be calculated. The work of \citet{Moreels2015b} utilises a uniform flux tube model, and calculates the wave energy by integrating over the entire flux cylinder and its exterior, before averaging over the period and wavelength of the oscillation, with the equations used in this case being derived for surface mode waves. Therefore, the resulting energies depend on the physical features of the pore (size, magnetic field strength, Alfv\'{e}n and sound speeds, and density), in addition to the properties intrinsic to the wave (period, phase speed and the Lagrangian displacement at the flux tube boundary). For further details, a full and detailed description of the theory required to calculate the energy of the waves is presented by \citet{Moreels2015b}.

The energy flux values for each wave are detailed in Table~1, with the energies plotted in Figure~{\ref{energy}} to highlight the observed dissipation with atmospheric height. It is seen that the waves initially contain in excess of $35{\,}000$~W{\,}m$^{-2}$ at the solar surface. The estimated wave energy decreases with height, with their associated energies at the {\cak} level being three orders-of-magnitude less than at their formation at the base of the magnetic pore. 

These results confirm the suitability of a magnetic pore to carry compressive wave energy into the chromosphere. The observed energy damping with height is directly related to the reduction in observed wave power at higher atmospheric heights. This may have physical meaning since the structuring of the pore is more noticeably diffuse at upper-chromospheric heights, hence suggesting the pore becomes unsuitable for efficient wave energy transportation. It is also expected that wave power will be lost due to the reflection of a portion of the waves. Our phase angle histograms suggest that approximately 30\% of identified wave activity between individual bandpasses propagates downwards. This can lead to a reduction in the observed wave energetics at higher atmospheric heights and, in part, explain the drop in energy flux from the photosphere to the chromosphere.

An explanation of the sharp drop in energy flux with atmospheric height could be attributed to leaky wave characteristics, i.e., waves that are freely able to release their energy into the surrounding plasma \citep{Cally1986}. Sausage mode oscillations displaying a leaky nature have already been observed by \citet{Mor2012}, and our suggestion for this is based upon the relatively long wavelengths of the observed oscillations. The wavelength of the oscillations can be deduced using a method presented by \citet{Mol1997}, and later applied by \citet{Zaq2007}, where the wavelength can be defined as,
\begin{equation}
\lambda = \frac{2\pi \delta s}{\delta \varphi} \ ,
\end{equation}
where $\delta s$ is the distance traversed by the waves and $\delta \varphi$ is the phase difference of the wave observed between the two distinct heights. However, when employing this equation one must consider a potential $2\pi$ uncertainty in the phase angle calculated from the phase difference analysis. Such uncertainties are mitigated here as a result of high cadence observations that allow us to isolate the start/end points of the oscillation cycles in each of the respective bandpasses. As a result, we have confidence that wave activity is observed first in filtergrams corresponding to the lower portions of the solar atmosphere, hence validating our interpretation that the waves are upwardly propagating. Using the values stated in Section~\ref{int} corresponding to the 4170{\,}{\AA} continuum to G-band transition, where $\delta s \sim 75$~km and $\delta \varphi = 6.12{\degr}$, the wavelength of the sausage modes can be estimated as $\lambda\approx4400$~km. This far exceeds the distance over which these waves are observed to propagate ($\sim1000$~km). Therefore, it appears that the observed drop in energy flux occurs within one quarter of the wavelength of the waves. Many currently proposed damping mechanisms hinge on the assumption that multiple cycles of the wave are present. As a result, none of these would adequately describe the wave damping we observe in the magnetic pore. 

An alternative scenario for the damping of the observed energy may be through mode conversion. As the waves propagate upwards, they will pass through a region where the plasma $\beta$ (i.e., the ratio of the plasma pressure to the magnetic pressure) equals $1$. In this regime favourable conditions exist for waves to couple and mode convert \citep[e.g.,][]{Ulm1991}. For magnetic regions of the Sun's atmosphere, the $\beta = 1$ layer is expected to be beneath the typical heights sampled by the {\cak} filter \citep{Jess12c}, thus mode conversion of slow sausage modes to fast modes within inclined magnetic fields may occur \citep{Bogdan2003}. Such mode conversion would likely occur at the inclined edges of the pore perimeter, and coupled with the propensity for fast modes to be leaky in nature, would help facilitate the rapid dissipation of wave energy into the surrounding plasma.

We must draw attention to the previously described model limitations, alongside other features that contribute uncertainties to the calculated energies. As mentioned in Section~\ref{waveperturb}, the model parameters used to describe the pore contain inaccuracies at different heights in the atmosphere, which will have an effect on the derived energy flux values. As mentioned previously, the magnetic field strengths used may be overestimated at photospheric heights, which will cause a subsequent overestimation of the wave energy flux in the 4170{\,}{\AA} continuum and G-band data sets. Alongside this, the potential underestimation of the pore density at chromospheric heights (i.e., through potentially overinflated expansion factors) may lead to the energy content of the waves in the {\nad} and {\cak} bandpasses being underestimated. Furthermore, the uniform flux tube model employed here has several limitations, i.e., it does not consider several physical effects such as density stratification, flux tube expansion and/or dissipation. However, while the uniform flux tube model does not consider these effects, it is still valid when used as a first order approximation, especially considering the pore studied here exhibits minimal expansion with atmospheric height. Thus, flux tube expansion will not have a significant impact on the overall energy flux since the $15$\% area expansion observed is relatively small. \citet{Andries2011} have shown that, for a slowly expanding flux tube, the perpendicular eigenfunctions \citep[which are used here to calculate the energy flux;][]{Moreels2015b} of MHD waves remain almost constant. In addition, density stratification is not accounted for in the equilibrium model to calculate the energy flux. On the other hand, the energy flux is calculated at each localised height separately using the equilibrium density at that height. Thus, density stratification is to some degree naturally included in the energy flux calculations. 

For future studies, combining a more realistic pore model atmosphere alongside a more detailed flux tube model may subsequently result in improved energy and dissipation estimates, particularly for atmospheric heights corresponding to the {\nad} and {\cak} image sequences. Furthermore, for display purposes in Figure~{\ref{energy}} we utilised the {\nad} and {\cak} formation heights estimated by \citet{Simon08} and \citet{Beebe1969}, which correspond to values approximately equal to $400$~km and $800$~km, respectively. As discussed above, these formation heights are relatively uncertain due to the passbands used sampling portions of the spectral line wings, in addition to changes in opacity across the magnetically structured pore itself. However, it is still observationally evident from Figure~{\ref{full}} that each bandpass samples a different region of the solar atmosphere, and as a result, still highlights the clear reduction in wave energy with atmospheric height even if the specific heights have a degree of uncertainty.

\subsection{Oscillations in the Magnetic Field}
In this paper, oscillations in the pore area, intensity and line-of-sight velocity have been utilised to investigate the characteristics of sausage mode waves. However, \citet{Fujimura2009} revealed that simultaneous perturbations in the associated magnetic flux should also be evident. The authors undertook analyses of the spectropolarimetric properties of the photospheric Fe~{\sc{i}}~$6302.5${\,}{\AA} absorption line, from data obtained by the Hinode spacecraft, to detect root mean squared oscillation amplitudes between $0.3$\% and $1.2$\% above the mean. Therefore, an additional investigation of fluctuations in the magnetic field, alongside those already considered, would provide further information about the mode and behaviour of the sausage waves, including the ability to diagnose non-trivial situations such as the  superposition of both fast and slow modes within the same structure \citep{Mor2013a}. Unfortunately, the data presented here does not have complimentary high-resolution spectropolarimetric observations. However, a natural question arises as to whether it is possible to detect magnetic field oscillations, particularly for larger-scale structures including solar pores, with more coarse resolution instruments such as HMI. Thus, the first important step is to calculate what magnetic field amplitude would be expected from the sausage mode oscillations detected in our present data set, and then compare these to the resolution and sensitivity of the HMI instrument. 

The magnetic flux, $\Phi$, passing through a surface is given by,
\begin{equation}
\Phi=\bf{B}\cdot\bf{S} \ ,
\end{equation}
where $\bf{B}$ and $\bf{S}$ are the vector magnetic field strength and pore area, respectively. The vector area is defined as, 
\begin{equation}
{\bf{S}}=\sum_i{\hat{{\bf{n}_i}}} {S_i} \ ,
\end{equation}
where $\hat{\bf{n}}$ and $S_i$ are the normal to the pore area and scalar area of the $i$th element of the pore surface, respectively. Now let us consider a surface with a normal vector in the direction of the magnetic field (i.e., ${\bf{B}} =B_0\hat{\bf{z}}=B_{z}$), providing a pore cross-section that is trapped in the $x$--$y$ plane with a cross-sectional area, $A$, so that $\hat{\bf{n}}S=\hat{\bf{z}}A$. Perturbing the variables, linearising, and assuming the net flux remains constant through the surface defined by the cross-sectional area, gives, 
\begin{equation}
\Phi=\bf{B}_0\cdot\bf{S}_0+\bf{B}_1\cdot\bf{S}_0+\bf{B}_0\cdot\bf{S}_1 \ ,
\end{equation}
or equivalently,
\begin{equation}
0={b}_z{A}_0+{B}_0{A}_1 \ ,
\end{equation}
where $A_1$ and $b_{z}$ are the perturbations in the area and the vertical (i.e., $z$ direction) magnetic field strength, respectively, with $A_0$ and $B_0$ denoting the mean unperturbed area and magnetic field values. Perhaps more importantly, we can re-write this relationship as,
\begin{equation}
\frac{A_1}{A_0}=-\frac{b_{z}}{B_0} \ .
\end{equation}
Hence, to be able to calculate the magnitude of the change in magnetic field strength induced as a result of the upwardly propagating sausage-mode waves, we need to utilise the area amplitudes derived for each observational bandpass that are detailed in Table~1.

Considering a photospheric average vertical magnetic field strength of the pore equal to approximately 900{\,}G (see, e.g., the lower-left panel of Figure~{\ref{full}}), and using the cross-sectional area and amplitude values detailed in Table~1, we are able to compute the magnitude of $b_z$ expected. Given the estimated formation heights for each bandpass, defined in section \ref{int}, the scaled \citet{Mal1986} magnetic field strengths corresponding to these heights can be selected (solid line in the lower-left panel of Figure~\ref{pore_parameters}). Thus, we can estimate the magnetic field oscillation amplitudes present in each bandpass to be $60${\,}G, $30${\,}G, $10${\,}G and $5${\,}G, for the 4170{\,}{\AA} continuum, G-band, {\nad} and {\cak} channels, respectively. For the HMI magnetogram formation height of $\sim$$300${\,}km, and using a $\approx$$2.2$\% area amplitude similar to that of G-band and {\nad} observations, we estimate that a magnetic field oscillation equal to $19${\,}G should be present in the HMI magnetograms. Such an oscillation with a peak amplitude $\approx$$19${\,}G is close to the noise limit inherent to the HMI instrument, where previous analyses by \citet{Liu2012} and \citet{Welsch2012} have detected uncertainties in the measured field strengths to be $\approx$$10${\,}Mx{\,}cm$^{-2}$ and $\approx$$30${\,}G, respectively. Furthermore, the cadence of the HMI vector ($720${\,}s) and line-of-sight ($45${\,}s) magnetograms are not ideally suited to accurately detect waves with periodicities as short as $3$~minutes, suggesting the reliable detection of magnetic field oscillations in pores is unlikely with the HMI instrument. However, exploiting this relationship in the future will allow comparisons to be made between observed changes in magnetic isosurfaces and the oscillations found within Stokes profiles \citep[e.g.,][]{Bal2000}. In addition, as there are known difficulties interpreting complex Stokes measurements, observing the changes in pore area could provide a novel way of validating the magnetic oscillations derived from simultaneous Stokes profiles. 

\section{Conclusion}
Observations of sausage mode waves propagating from the solar photosphere to the base of the transition region are presented here for the first time. High resolution, multi-wavelength datasets of a magnetic pore are investigated using wavelet and Fourier analyses to identify oscillations within the pore structure. Simultaneous oscillations in both intensity emission and the area encompassed by the pore provide evidence of slow sausage mode activity. Through analysis of these oscillations in adjoining bandpasses, it is identified that sausage modes with periods between $181-412$~s are predominantly propagating upwards from the photosphere to the chromosphere. The phase angles calculated for these waves between adjacent bandpasses indicate a phase speed in excess of $15$~km{\,}s$^{-1}$. However, the line-of-sight velocity information for this pore shows a strong simultaneous plasma upflow, which is believed to dominate these high speeds. As a result, the initial velocity calculation can be considered as the Doppler-shifted phase speed of the waves. Using theoretical considerations, the waves are verified to be slow surface modes, propagating with a phase speed on the order of $3$~km{\,}s$^{-1}$. The observed periods of the waves are similar to that of the global solar acoustic oscillations, and provide further insight that sausage mode waves may be driven by the ubiquitous $p$-modes. 

For the first time, a multi-layer analyses of the energy content of the sausage mode waves is documented. Individual sausage modes were identified for further study, with their specific oscillation amplitudes in intensity and area deduced. It can be seen in Table~\ref{Periods} that the oscillation amplitudes increased between photospheric and chromospheric heights, consistent with what would be expected as the waves propagate into less dense regions, since there will be less external restoring forces for the area perturbations. Through the identification of these amplitudes, the energy flux of the waves at each bandpass was calculated. It is seen that the waves are formed at the solar surface with a large energy magnitude in excess of $35{\,}000$~W{\,}m$^{-2}$, which significantly reduces as they propagate into the chromosphere. The limitations of the energy flux model are also presented, and we suggest how improved atmospheric and flux tube models may be able to help more accurately constrain the calculated wave energies as a function of atmospheric height. However, importantly we show that magnetic pores can act as viable conduits for sausage mode waves to transport energy to higher atmospheric heights. The energy flux values calculated show that these waves have lost the vast majority of their energy before they reach the chromospheric layer observed in {\cak} filtergrams. This could be due to wave reflection and conventional damping methods, which are difficult to identify observationally, or that these waves are inherent leaky modes and freely able to radiate energy into the surroundings. However, MHD waves previously observed in chromospheric structures have exhibited far greater localised energy flux, with Alfv{\'{e}}nic waves identified to contain $4-7$~kW{\,}m$^{-2}$ \citep{DePontieu2007}, Alfv{\'{e}}n waves with energy flux of ~$15$~kW{\,}m$^{-2}$ \citep{Jess2009} and sausage modes displaying approximately $11{\,}700$~W{\,}m$^{-2}$ \citep{Mor2012}. The smaller chromospheric energy values detected in our observations, compared to previous studies, could be a result of the structuring of the pore itself. The magnetic pore under consideration is small and dynamic, is therefore more difficult to resolve from the background in chromospheric images, and cannot be positively identified in simultaneous higher atmospheric imaging by SDO. The lack of pore structuring at higher heights may be an indication of it's instability, and thus the pore is unable to remain a viable energy conduit for compressible MHD waves at these higher atmospheric heights. 

Due to any proposed heating occuring below observable MHD scales, there is a lack of observational evidence for the heating of chromospheric plasma as a result of the damping of sausage mode energy seen in the pore. Instead, the results presented here help to clarify the validity of magnetic pores as energy conduits for sausage mode waves. Further investigation of sausage modes in a variety of pores at high spatial, temporal and spectral resolutions may provide a clearer picture on the energy content and dissipative potential of these waves. With the advent of the Interface Region Imaging Spectrograph \citep[IRIS;][]{DeP14} mission, and the unprecedented resolution capabilities of the upcoming Daniel K. Inouye Solar Telescope \citep[DKIST, formerly the Advanced Technology Solar Telescope, ATST;][]{Kei03, Rim10}, there will be new and novel ways to probe the importance that magnetic pores and sausage mode waves play in the heating of both the chromosphere and corona.  

\acknowledgments
We thank the anonymous referee for their valuable input into this manuscript. S.D.T.G. thanks the Northern Ireland Department for Employment and Learning for the award of a Ph.D. studentship. D.B.J. wishes to thank the UK Science and Technology Facilities Council for the award of an Ernest Rutherford Fellowship alongside a dedicated Research Grant. M.G.M. and T.V.D. have received funding from the Odysseus programme of the FWO-Vlaanderen. The research was conducted in the framework of Belspo's IAP P7/08 CHARM and the GOA-2015-014 of the Research Council of the KU Leuven. D.J.C. thanks the CSUN Department of Physics and Astronomy for their support of CSUNcam. G.V. acknowledges the Leverhulme Trust (UK). R.E. is grateful to NSF, Hungary (OTKA, Ref. No. K83133) and acknowledges the financial support recieved from the Science and Technology Facilities Council (STFC), UK

\end{document}